\documentclass{llncs}
\usepackage[utf8]{inputenc}
\usepackage[T1]{fontenc}
\usepackage{amsfonts}

\newcommand{\packageGraphicx}{\usepackage{graphicx}}
\newcommand{\packageHyperref}{\usepackage{hyperref}}
\newcommand{\renewrmdefault}{\renewcommand{\rmdefault}{ptm}}
\newcommand{\packageRelsize}{\usepackage{relsize}}
\newcommand{\packageMathabx}{\usepackage{mathabx}}
\newcommand{\packageWasysym}{
  \let\leftmoon\relax \let\rightmoon\relax \let\fullmoon\relax \let\newmoon\relax \let\diameter\relax
  \usepackage{wasysym}}
\newcommand{\packageTextcomp}{\usepackage{textcomp}}
\newcommand{\packageFramed}{\usepackage{framed}}
\newcommand{\packageHyphenat}{\usepackage[htt]{hyphenat}}
\newcommand{\packageColor}{\usepackage[usenames,dvipsnames]{color}}
\newcommand{\doHypersetup}{\hypersetup{bookmarks=true,bookmarksopen=true,bookmarksnumbered=true}}
\newcommand{\packageTocstyle}{\IfFileExists{tocstyle.sty}{\usepackage{tocstyle}\usetocstyle{standard}}{}}
\newcommand{\packageCJK}{\IfFileExists{CJK.sty}{\usepackage{CJK}}{}}

\packageGraphicx
\packageHyperref
\renewrmdefault
\packageRelsize
\packageMathabx
\packageWasysym
\packageTextcomp
\packageFramed
\packageHyphenat
\packageColor
\doHypersetup
\packageTocstyle
\packageCJK

\newcommand{\sectionNewpage}{}
\newcommand{\preDoc}{}
\newcommand{\postDoc}{}

\newcommand{\BookRef}[2]{\emph{#2}}
\newcommand{\ChapRef}[2]{\SecRef{#1}{#2}}
\newcommand{\SecRef}[2]{section~#1}
\newcommand{\PartRef}[2]{part~#1}
\newcommand{\BookRefUC}[2]{\BookRef{#1}{#2}}
\newcommand{\ChapRefUC}[2]{\SecRefUC{#1}{#2}}
\newcommand{\SecRefUC}[2]{Section~#1}
\newcommand{\PartRefUC}[2]{Part~#1}

\newcommand{\BookRefLocal}[3]{\hyperref[#1]{\BookRef{#2}{#3}}}
\newcommand{\ChapRefLocal}[3]{\hyperref[#1]{\ChapRef{#2}{#3}}}
\newcommand{\SecRefLocal}[3]{\hyperref[#1]{\SecRef{#2}{#3}}}
\newcommand{\PartRefLocal}[3]{\hyperref[#1]{\PartRef{#2}{#3}}}
\newcommand{\BookRefLocalUC}[3]{\hyperref[#1]{\BookRefUC{#2}{#3}}}
\newcommand{\ChapRefLocalUC}[3]{\hyperref[#1]{\ChapRefUC{#2}{#3}}}
\newcommand{\SecRefLocalUC}[3]{\hyperref[#1]{\SecRefUC{#2}{#3}}}
\newcommand{\PartRefLocalUC}[3]{\hyperref[#1]{\PartRefUC{#2}{#3}}}

\newcommand{\BookRefUN}[1]{\BookRef{}{#1}}

\newcommand{\SecRefUN}[1]{``#1''}

\newcommand{\BookRefLocalUN}[2]{\hyperref[#1]{\BookRefUN{#2}}}

\newcommand{\SecRefLocalUN}[2]{\hyperref[#1]{\SecRefUN{#2}}}

\newcommand{\SectionNumberLink}[2]{\hyperref[#1]{#2}}

\newcommand{\Scribtexttt}[1]{{\texttt{#1}}}

\newcommand{\planetName}[1]{PLane\hspace{-0.1ex}T}

\newcommand{\Stttextless}{{\fontencoding{T1}\selectfont<}}
\newcommand{\Stttextbar}{{\fontencoding{T1}\selectfont|}}

\makeatletter
\def\empty@finalstrut#1{%
  \unskip\ifhmode\nobreak\fi\vrule\@width\z@\@height\z@\@depth\z@}
\def\no@strut{\global\setbox\@arstrutbox\hbox{%
    \vrule \@height\z@
           \@depth\z@
           \@width\z@}%
    \gdef\@endpbox{\empty@finalstrut\@arstrutbox\par\egroup\hfil}%
}%
\def\yes@strut{\global\setbox\@arstrutbox\hbox{%
    \vrule \@height\arraystretch \ht\strutbox
           \@depth\arraystretch \dp\strutbox
           \@width\z@}%
    \gdef\@endpbox{\@finalstrut\@arstrutbox\par\egroup\hfil}%
}%
\def\@mkpream#1{\@firstamptrue\@lastchclass6
  \let\@preamble\@empty\def\empty@preamble{\add@ins}%
  \let\protect\@unexpandable@protect
  \let\@sharp\relax\let\add@ins\relax
  \let\@startpbox\relax\let\@endpbox\relax
  \@expast{#1}%
  \expandafter\@tfor \expandafter
    \@nextchar \expandafter:\expandafter=\reserved@a\do
       {\@testpach\@nextchar
    \ifcase \@chclass \@classz \or \@classi \or \@classii \or \@classiii
      \or \@classiv \or\@classv \fi\@lastchclass\@chclass}%
  \ifcase \@lastchclass \@acol
      \or \or \@preamerr \@ne\or \@preamerr \tw@\or \or \@acol \fi}
\def\@addamp{%
  \if@firstamp
    \@firstampfalse
    \edef\empty@preamble{\add@ins}%
  \else
    \edef\@preamble{\@preamble &}%
    \edef\empty@preamble{\expandafter\noexpand\empty@preamble &\add@ins}%
  \fi}
\newif\iftw@hlines \tw@hlinesfalse
\def\@xhline{\ifx\reserved@a\hline
               \tw@hlinestrue
             \else\ifx\reserved@a\Hline
               \tw@hlinestrue
             \else
               \tw@hlinesfalse
             \fi\fi
      \iftw@hlines
        \aftergroup\do@after
      \fi
      \ifnum0=`{\fi}%
}
\def\do@after{\emptyrow[\the\doublerulesep]}
\def\emptyrow{\noalign\bgroup\@ifnextchar[\@emptyrow{\@emptyrow[\z@]}}
\def\@emptyrow[#1]{\no@strut\gdef\add@ins{\vrule \@height\z@ \@depth#1 \@width\z@}\egroup%
\empty@preamble\\
\noalign{\yes@strut\gdef\add@ins{\vrule \@height\z@ \@depth\z@ \@width\z@}}%
}
\def\tabrow#1{\noalign\bgroup\@ifnextchar[{\@tabrow{#1}}{\@tabrow{#1}[]}}
\def\@tabrow#1[#2]{\no@strut\egroup#1\ifx.#2.\\\else\\[#2]\fi\noalign{\yes@strut}}
\def\endpltstabular{\crcr\egroup\egroup \egroup}
\expandafter \let \csname endpltstabular*\endcsname = \endpltstabular
\def\pltstabular{\let\@halignto\@empty\@pltstabular}
\@namedef{pltstabular*}#1{\def\@halignto{to#1}\@pltstabular}
\def\@pltstabular{\leavevmode \bgroup \let\@acol\@tabacol
   \let\@classz\@tabclassz
   \let\@classiv\@tabclassiv \let\\\@tabularcr\@stabarray}
\def\@stabarray{\m@th\@ifnextchar[\@sarray{\@sarray[c]}}
\def\@sarray[#1]#2{%
  \bgroup
  \setbox\@arstrutbox\hbox{%
    \vrule \@height\arraystretch\ht\strutbox
           \@depth\arraystretch \dp\strutbox
           \@width\z@}%
  \@mkpream{#2}%
  \edef\@preamble{%
    \ialign \noexpand\@halignto
      \bgroup \@arstrut \@preamble \tabskip\z@skip \cr}%
  \let\@startpbox\@@startpbox \let\@endpbox\@@endpbox
  \let\tabularnewline\\%
    \let\@sharp##%
    \set@typeset@protect
    \lineskip\z@skip\baselineskip\z@skip
    \@preamble}

\makeatother

\newlength{\stabLeft}

\newcommand{\atItemizeStart}[0]{\addtolength{\stabLeft}{\labelsep}
                                \addtolength{\stabLeft}{\labelwidth}}

\newenvironment{SingleColumn}{\begin{list}{}{\topsep=0pt\partopsep=0pt%
\listparindent=0pt\itemindent=0pt\labelwidth=0pt\leftmargin=0pt\rightmargin=0pt%
\itemsep=0pt\parsep=0pt}\item}{\end{list}}

\newcommand{\SCodePreSkip}{\vskip\abovedisplayskip}
\newcommand{\SCodePostSkip}{\vskip\belowdisplayskip}

\newcommand{\SVInsetPreSkip}{\vskip\abovedisplayskip}
\newcommand{\SVInsetPostSkip}{\vskip\belowdisplayskip}

\newcommand{\titleAndVersionAndAuthors}[3]{\title{#1\\{\normalsize \SVersionBefore{}#2}}\author{#3}\maketitle}

\newcommand{\titleAndEmptyVersionAndAuthors}[3]{\title{#1}\author{#3}\maketitle}

\newcommand{\SAuthor}[1]{#1}
\newcommand{\SAuthorSep}[1]{\qquad}
\newcommand{\SVersionBefore}[1]{Version }

\newcommand{\SNumberOfAuthors}[1]{}

\let\SOriginalthesubsection\thesubsection
\let\SOriginalthesubsubsection\thesubsubsection

\newcommand{\Ssection}[2]{\section[#1]{#2}\let\thesubsection\SOriginalthesubsection}
\newcommand{\Ssubsection}[2]{\subsection[#1]{#2}\let\thesubsubsection\SOriginalthesubsubsection}

\newcounter{GrouperTemp}

\newcommand{\Snolinkurl}[1]{\nolinkurl{#1}}

\newcommand{\LNCSand}{\and}

\newcommand{\LNCSauthor}[1]{\author{#1}}

\newcommand{\LNCSinstitutes}[1]{\institute{#1}}

\newcommand{\LNCSemail}[1]{\email{#1}}

\renewcommand{\SAuthorSep}{}

\def\texMathInline#1{\ensuremath{#1}}
\def\texMathDisplay#1{\ifmmode #1\else\[#1\]\fi}
\newcommand{\NoteBox}[1]{\footnote{#1}}
\newcommand{\NoteContent}[1]{#1}

\newcommand{\FootnoteRef}[1]{}
\newcommand{\FootnoteTarget}[1]{}

\newcommand{\FootnoteBlockContent}[1]{}
\usepackage{ccaption}

\makeatletter
\@ifundefined{belowcaptionskip}{}{}
\makeatother

\newcommand{\Legend}[1]{~

                        \hrule width \hsize height .33pt
                        \vspace{4pt}
                        \legend{#1}}

\newcommand{\FigureTarget}[2]{#1}
\newcommand{\FigureRef}[2]{#1}

\newlength{\FigOrigskip}
\FigOrigskip=\parskip

\newcommand{\FigureSetRef}{\refstepcounter{figure}}

\newenvironment{FigureMulti}{\begin{figure*}[t!p]\FigureSetRef}{\end{figure*}}

\newenvironment{Herefigure}{\begin{figure}[ht!]\FigureSetRef\centering}{\end{figure}}

\newenvironment{Centerfigure}{\begin{Xfigure}\centering\item}{\end{Xfigure}}

\newenvironment{Xfigure}{\begin{list}{}{\leftmargin=0pt\topsep=0pt\parsep=\FigOrigskip\partopsep=0pt}}{\end{list}}

\newenvironment{FigureInside}{}{}

\newcommand{\Centertext}[1]{\begin{center}#1\end{center}}

\usepackage{calc}
\newlength{\ABcollength}

\renewcommand{\titleAndVersionAndAuthors}[3]{\title{#1}\titlerunning{#2}#3\maketitle}
\renewcommand{\titleAndEmptyVersionAndAuthors}[3]{\titleAndVersionAndAuthors{#1}{#1}{#3}}
\begin{document}
\preDoc
\titleAndEmptyVersionAndAuthors{Compiling Control as Offline Partial Deduction}{}{\SNumberOfAuthors{2}\SAuthor{\LNCSauthor{Vincent Nys\LNCSand{}Danny De Schreye}}\SAuthorSep{}\SAuthor{\LNCSinstitutes{KU Leuven\hspace*{\fill}\\\LNCSemail{\{vincent.nys,danny.deschreye\}@kuleuven.be}}}}
\label{t:x28part_x22Compilingx5fControlx5fasx5fOfflinex5fPartialx5fDeductionx22x29}

\noindent 

\begin{abstract}We present a new approach to a technique known as compiling control, whose aim is to compile away special mechanisms for non{-}standard atom selection in logic programs. It has previously been conjectured that compiling control could be implemented as an instance of the first Futamura projection, in which an interpreter is specialized for an input program. However, the exact nature of such an interpreter and of the required technique for specialization were never specified. In this work, we propose a Prolog meta{-}interpreter which applies the desired non{-}standard selection rule and which is amenable to specialization using offline partial deduction. After the initial analysis phase of compiling control, we collect annotations to specialize the interpreter using the Logen system for offline partial deduction. We also show that the result of the specialization is equivalent to the program obtained using the traditional approach to compiling control. In this way, we simplify the synthesis step.\end{abstract}


\sectionNewpage

\Ssection{Introduction}{Introduction}\label{t:x28part_x22Introductionx22x29}

Compiling control is a program transformation technique which aims to compile the runtime behavior of pure logic programs executed under a non{-}standard selection rule to logic programs which are totally equivalent under the standard, left{-}to{-}right selection rule of Prolog. It was originally presented in~\cite{bruynooghe_1986_compiling} and~\cite{bruynooghe_1989_compiling}.
The technique is designed to work in two phases. In a first phase, the computation flow of the program, executed under the non{-}standard rule, is analyzed, resulting in a symbolic evaluation tree that captures the entire flow. In a second phase, from the symbolic evaluation tree, a new logic program is synthesized. The technique was formalized and proven correct under certain technical conditions, but it possessed certain drawbacks. Most importantly, it was an ad hoc solution. Because of this, showing that the analysis phase of the transformation was complete for a specific program required a manual proof by induction.
However, since the original presentation of compiling control, several frameworks have been developed which provide a more formal perspective on the two phases of compiling control and whose general correctness results can be reused. Most notably, abstract interpretation~\cite{bruynooghe_1991_practical} and partial deduction~\cite{komorowski_1981_specification}~\cite{gallagher_1986_transforming}.
In~\cite{nys_2017_abstract}, we showed that the technique could in some cases be reformulated and formalized using abstract conjunctive partial deduction, a framework proposed in~\cite{leuschel_2004_framework} which integrates the aforementioned frameworks. In addition, we proposed a new abstraction, \texMathInline{multi}, to analyze computations with unboundedly growing goals. This allowed us to analyze a diverse set of well{-}known programs and to compile these into programs executed under the standard selection rule. Unfortunately, the \texMathInline{multi} abstraction also broke an explicit assumption of the abstract conjunctive partial deduction framework.
In the current work, we propose a different perspective. We show that the synthesis obtained using the previously published approach can also be obtained by applying the first Futamura projection, in which program specialization is applied to an interpreter and an input program. Such an approach has been speculated upon in the past, but the current work is the first that demonstrates that this is indeed feasible. It is not an instance of the abstract conjunctive partial deduction framework, but rather a standard offline partial deduction, which implies that no changes to the abstract conjunctive partial deduction framework are necessary to relax the aforementioned assumption which is not met. Based on our experiments, there are no programs which can be compiled using the classical approach but not the approach presented here.

The idea of applying the first Futamura projection to obtain a more structured representation of control flow can also be found in~\cite{zamalloa_2009_decompilation}, where Java bytecode is decompiled to Prolog.
The notion of modelling and analyzing the execution of a (PIC) program as a logic program and partially evaluating that can be found in~\cite{henriksen_2006_abstract}.
This is akin to what we do, though we abstractly analyze the program to be executed itself, which is already a logic program, and partially evaluate the interpreter.
Other examples of specialization of logic program interpreters, also using the Logen system, are provided in~\cite{leuschel_2004_specialising}.

We will first give a brief introduction to the first Futamura projection and to offline partial deduction in \ChapRefLocalUC{t:x28part_x22preliminariesx22x29}{2}{Preliminaries}.
Then, we will provide a motivating example.
We will explain the notation and operations shown in the motivating example by introducing our abstract domain in \ChapRefLocalUC{t:x28part_x22abstractx2ddomainx22x29}{4}{Abstract domain}.
We will use the abstract domain to express the scope of the technique in \ChapRefLocalUC{t:x28part_x22scopex22x29}{5}{Instantiation}.
Next, in \ChapRefLocalUC{t:x28part_x22analysisx2dphasex22x29}{6}{The analysis phase}, we will round out the analysis for the first example program.
We will also show a simple meta{-}interpreter which, using information obtained from the analysis, can be configured to implement the desired non{-}standard selection rule.
Once the basic idea has been illustrated, we will show the most interesting parts of the analysis of a more complex program, as well as extensions to the meta{-}interpreter required to run this program in a  satisfactory way in \ChapRefLocalUC{t:x28part_x22primesx22x29}{8}{Programs requiring the \texMathInline{multi} abstraction: primes}.
In \ChapRefLocalUC{t:x28part_x22specializationx22x29}{9}{Specialization using Logen}, we will show how the meta{-}interpreter can be specialized using the Logen system for offline partial deduction.
Then, in \ChapRefLocalUC{t:x28part_x22synthesisx2dequivalencex22x29}{10}{Equivalence with the classical approach}, we will show that the obtained specialization is indeed equivalent to the "classical" synthesis.
We wrap up with a discussion and with avenues for future work.
A set of example programs along with corresponding analyses and syntheses using both techniques is available as an electronic appendix at \href{https://perswww.kuleuven.be/~u0055408/cc-as-opd.html}{https{\hbox{\texttt{:}}}//perswww{\hbox{\texttt{.}}}kuleuven{\hbox{\texttt{.}}}be/$\sim$u0055408/cc{-}as{-}opd{\hbox{\texttt{.}}}html}.

\sectionNewpage

\Ssection{Preliminaries}{Preliminaries}\label{t:x28part_x22preliminariesx22x29}

In~\cite{futamura_1999_partial}, Futamura showed that partially evaluating an interpreter for a language \texMathInline{l_1} (written in language \texMathInline{l_2}) for a "source program" (in \texMathInline{l_1}) yields an "object program" (in \texMathInline{l_2}) with the semantics of the source program, as run by the interpreter. That is, partially evaluating an interpreter for a source program is an act of compiling.

This is expressed by the following equation: \texMathInline{int(s,r)=\alpha(int,s)(r)}. Here, \texMathInline{int} is the function encoded by the interpreter. Its arguments \texMathInline{s} and \texMathInline{r} are the "static" and "runtime" inputs, i.e. inputs which remain constant and which may vary between executions, respectively. In the setting of partial evaluation of an interpreter, the "static" input is the program to be interpreted, whereas the "runtime" input is the input to the interpreted program.
The function \texMathInline{\alpha} is the partial evaluation function: it specializes its first argument for the static input. In this setting, this produces a program which takes the runtime inputs, so \texMathInline{r}, and behaves as the interpreter would when also given \texMathInline{s}. Therefore \texMathInline{\alpha(int,s)} is the compiled version of the source program which Futamura calls the "object program". The observation that a compiled program can be obtained through specialization of an interpreter is known as the first Futamura projection~\cite{jones_1993_partial}.

Partial deduction is a technique for logic program specialization originally introduced in~\cite{komorowski_1981_specification} and formalized in~\cite{lloyd_1991_partial}.
The idea behind partial deduction is to compute a set of derivation trees for some top level goal \texMathInline{A} such that all expected queries instantiate \texMathInline{A}.
The computations represented by the branches of these trees can then be collected into logical implications, referred to as "resultants".
These are then encoded as logic program clauses, so that a program is obtained which is equivalent to the original program for all queries which instantiate \texMathInline{A}, but not necessarily for other queries.

An important notion is "closedness".
Under closedness, all atoms in a partial deduction are instances of an atom in the root of a derivation tree.
This implies that a goal is never reduced to a goal that has not been specialized and that completeness is ensured.
Because the set of all roots of trees is denoted \texMathInline{\mathcal{A}}, this is also referred to as \texMathInline{\mathcal{A}}{-}closedness.

There are two broad approaches for dealing with the issue of control~\cite{leuschel_2002_logic}.
In "online partial deduction", control is tackled during the specialization phase itself.
That is, the construction of SLDNF{-}trees is monitored and unfolding continues as long as there is evidence that interesting computations are performed.

Offline partial deduction is different in that control decisions are taken before the specialization phase.
These are then cast in the form of program annotations for the specializer.
During the specialization phase, unfolding proceeds in a left{-}to{-}right fashion.
Broadly speaking, depending on the annotation of a call, it may be unfolded, it may be treated as an instance of a specialized atom or it may be kept residual.
Given the annotations, specialization is straightforward.
Annotations can be written manually, but can also be derived automatically in a separate phase.
This phase is referred to as a "binding{-}time analysis" and is performed before the static input is available.

The above concepts and techniques can be generalized to "conjunctive partial deduction".
In conjunctive partial deduction, the roots and leaves of derivation trees need not be atoms, but can also be conjunctions.
This adds a layer of complexity:
While partial deduction splits up conjunctions into atoms before starting new SLDNF{-}trees, conjunctive partial deduction has more options for splitting them into subconjunctions.

\sectionNewpage

\Ssection{Running example: permutation sort}{Running example: permutation sort}\label{t:x28part_x22runningx2dexamplex22x29}

To introduce the essential components of compiling control, we will use a motivating example.
The following is a Prolog implementation of permutation sort, which sorts a list by permuting it and then checking if the permutation is ordered correctly.

\begin{SingleColumn}\Scribtexttt{permsort(X,Y) {\hbox{\texttt{:}}}{-} perm(X,Y), ord(Y){\hbox{\texttt{.}}}}

\Scribtexttt{perm([],[]){\hbox{\texttt{.}}}}

\Scribtexttt{perm([X{\Stttextbar}Y],[U{\Stttextbar}V]) {\hbox{\texttt{:}}}{-} select(U,[X{\Stttextbar}Y],W), perm(W,V){\hbox{\texttt{.}}}}

\Scribtexttt{ord([]){\hbox{\texttt{.}}}}

\Scribtexttt{ord([X]){\hbox{\texttt{.}}}}

\Scribtexttt{ord([X,Y{\Stttextbar}Z]) {\hbox{\texttt{:}}}{-} X ={\Stttextless} Y, ord([Y{\Stttextbar}Z]){\hbox{\texttt{.}}}}\end{SingleColumn}

While this sorting program clearly expresses the declarative perspective on sorting as creating an ordered permutation, its naive implementation is problematic.
Its efficiency can be improved by using a different selection rule.
Informally, such a selection rule interleaves calls which build the permutation with calls which check for the correct ordering of elements.
Specifically, as soon as the first two elements of the permutation have been generated, their ordering can be checked.
\hyperref[t:x28counter_x28x22figurex22_x22permsort1x22x29x29]{Figure~\FigureRef{1}{t:x28counter_x28x22figurex22_x22permsort1x22x29x29}} and \hyperref[t:x28counter_x28x22figurex22_x22permsort2x22x29x29]{Figure~\FigureRef{2}{t:x28counter_x28x22figurex22_x22permsort2x22x29x29}} can be considered as symbolic derivation trees representing a computation under this more efficient selection rule.
In these figures, a value \texMathInline{a_i}, where \texMathInline{i} is a natural number, stands for any term, whereas a value \texMathInline{g_j}, where \texMathInline{j} is also a natural number, stands for a ground term.
When an atom is underlined once, we consider the effects of resolving said atom.
If it is underlined twice, we treat it as a built{-}in.
The reader is not intended to understand every aspect of the two trees at this point.
They are only intended to give an idea of the type of computation the synthesized program should execute.
In the following sections, we will provide formal underpinnings for the data and operations in these symbolic trees and will explain how they can be used to synthesize a program which simulates the program with the desired selection rule.

\begin{Herefigure}\begin{Centerfigure}\begin{FigureInside}\includegraphics[scale=1.0]{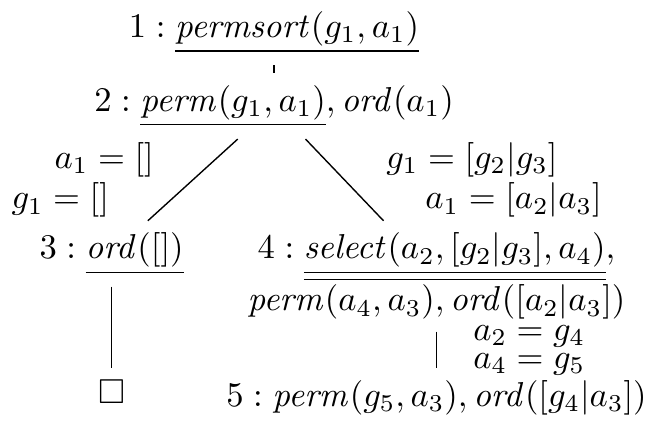}\end{FigureInside}\end{Centerfigure}

\Centertext{\Legend{\FigureTarget{\label{t:x28counter_x28x22figurex22_x22permsort1x22x29x29}Figure~1: }{t:x28counter_x28x22figurex22_x22permsort1x22x29x29}First analysis tree for permutation sort}}\end{Herefigure}

\begin{Herefigure}\begin{Centerfigure}\begin{FigureInside}\includegraphics[scale=1.0]{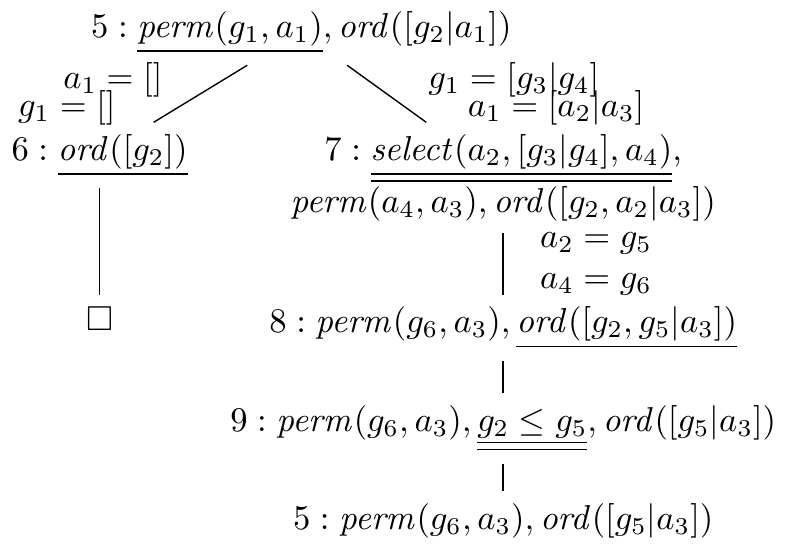}\end{FigureInside}\end{Centerfigure}

\Centertext{\Legend{\FigureTarget{\label{t:x28counter_x28x22figurex22_x22permsort2x22x29x29}Figure~2: }{t:x28counter_x28x22figurex22_x22permsort2x22x29x29}Second analysis tree for permutation sort}}\end{Herefigure}

\sectionNewpage

\Ssection{Abstract domain}{Abstract domain}\label{t:x28part_x22abstractx2ddomainx22x29}

The trees shown in \ChapRefLocalUC{t:x28part_x22runningx2dexamplex22x29}{3}{Running example: permutation sort} constitute the analysis phase of compiling control, which is a form of abstract interpretation.
This abstract interpretation remains entirely the same as in~\cite{nys_2017_abstract}.
We reproduce the key points here.

As any abstract interpretation, the analysis phase is based on an abstract domain, whose elements represent \textit{sets} of \textit{concrete} values with specific properties.
The fundamental building blocks of the abstract domain are two types of abstract variables, \texMathInline{a_i} and \texMathInline{g_j}.
An abstract variable \texMathInline{a_i} represents the set of all concrete values, whereas a variable \texMathInline{g_j} represents all ground concrete values.
The union of these two sets is denoted \texMathInline{AV\!ar_P}.

Abstract counterparts to concrete program constants are included in the abstract domain.
These represent the singleton sets consisting of the corresponding concrete constants.
This is why, in the example in \ChapRefLocalUC{t:x28part_x22runningx2dexamplex22x29}{3}{Running example: permutation sort}, the empty list \texMathInline{[]} occurs as an argument.
From these abstract variables and abstract constants, abstract terms, abstract atoms and abstract conjunctions are constructed, yielding the sets \texMathInline{AT\!erm_P}, \texMathInline{\mathit{A\!Atom}_P} and \texMathInline{AC\!on\!Atom_P}.
Example members of these sets are \texMathInline{[g_2|g_3]}, \texMathInline{permsort(g_1,a_1)} and \texMathInline{perm(g_1,a_1),ord(a_1)}, respectively.
If an abstract term, atom or conjunction contains some \texMathInline{a_i} or \texMathInline{g_j} several times (i.e. the occurrences have the same subscript), then the represented concrete term, atom or conjunction contains the \textit{same} subterm at every position corresponding to the positions of \texMathInline{a_i} or \texMathInline{g_j}.
For instance, in the second node in the running example, the result of the permutation operation, \texMathInline{a_1}, must be ordered.
Note that two abstract variables \texMathInline{a_i} and \texMathInline{g_j}, when \texMathInline{i = j}, are not assumed to be aliased.

Let \texMathInline{AT\!erm_{P/\approx}}, \texMathInline{A\!Atom_{P/\approx}} and \texMathInline{AC\!on\!Atom_{P/\approx}} denote the sets of equivalence classes of abstract terms, abstract atoms and abstract conjunctions, respectively.
Equivalence of abstract terms (or atoms or conjunctions) is based on abstract substitutions, which are finite sets of ordered pairs in \texMathInline{AV\!ar_P \times AT\!erm_P} and whose application instantiates abstract variables in a way that is analogous to how applying substitutions in the concrete domain instantiate concrete variables.
Two abstract terms (or atoms or conjunctions) \texMathInline{A} and \texMathInline{B} are equivalent, denoted \texMathInline{A \approx B}, if and only if there are abstract substitutions \texMathInline{\theta_1} and \texMathInline{\theta_2} such that \texMathInline{A\theta_1 = B} and \texMathInline{B\theta_2 = A}.

The abstract domain, \texMathInline{AD\!om_P}, is the union of \texMathInline{AV\!ar_P}, \texMathInline{AT\!erm_{P/\approx}} \texMathInline{A\!Atom_{P/\approx}} and \texMathInline{AC\!on\!Atom_{P/\approx}} (leaving aside the \texMathInline{multi} abstraction until \ChapRefLocalUC{t:x28part_x22primesx22x29}{8}{Programs requiring the \texMathInline{multi} abstraction: primes}).
Note that we will refer to an equivalence class by taking a representative.
For instance, we write \texMathInline{permsort(g_1,a_1)} when we mean "all abstract atoms equivalent to \texMathInline{permsort(g_1,a_1)}" and assume that the intended meaning is clear from the context.
Finally, let \texMathInline{D\!om_P} be the concrete domain and let \texMathInline{\gamma : AD\!om_P \rightarrow 2^{D\!om_P}} be the concretization function, which maps elements of the abstract domain to their concrete denotation.
For example, the denotation of \texMathInline{permsort(g_1,a_1)} is an infinite set of concrete atoms with predicate symbol \texMathInline{permsort}, with a ground first argument and any kind of second argument.

\sectionNewpage

\Ssection{Instantiation}{Instantiation}\label{t:x28part_x22scopex22x29}

In general, abstract interpretation requires a "widening" operator to achieve termination.
A widening operator replaces one abstract value with another, more general abstract value, which can come at the cost of accuracy of the analysis.
For our abstract domain, depth{-}\texMathInline{k} abstraction is such a type of widening.
Depth{-}\texMathInline{k} abstraction entails that any abstract values whose term depth exceeds a certain limit \texMathInline{k} are replaced with more general terms.
For instance, if only one level of term nesting is allowed and the atom \texMathInline{ord([g_1,g_2|a_1])} is computed, this term must be generalized.
The most specific term which does not exceed the nesting limit is \texMathInline{ord([g_1|a_2])}.
If such a widening were applied in the running example, the resulting synthesis would not simulate the desired selection rule.
In general, applying depth{-}\texMathInline{k} abstraction \emph{may} affect the correctness of the technique, depending on the program and the level of nesting which is allowed.
In what follows, we will assume that depth{-}\texMathInline{k} abstraction is not required for termination.

The abstract domain is also tied to the selection rule.
We assume that the selection rule is an instantiation based selection rule, which is defined as follows:

\relax{\vspace{1ex}
}\textbf{Definition} (\textit{instantiation{-}based selection rule}) \hspace*{\fill}\\An instantiation{-}based selection rule for \texMathInline{P} is a strict partial order \texMathInline{<} on \texMathInline{A\!Atom_{P/\approx}}, such that \texMathInline{\gamma(s_1) \subset \gamma(s_2)} implies \texMathInline{s_2 \nless s_1}, where \texMathInline{\subset} denotes strict set inclusion.\relax{\vspace{1ex}
}

An instantiation{-}based selection rule expresses which atom is selected from an abstract conjunction. Our technique requires that an instantiation{-}based selection rule can completely specify the desired control flow. This is formalized as follows.

\relax{\vspace{1ex}
}\textbf{Definition} (\textit{complete instantiation{-}based selection rule}) \hspace*{\fill}\\An instantiation{-}based selection rule for a program \texMathInline{P} is complete if, for each \texMathInline{A \in AC\!on_{P/\approx}}, there exists an abstract atom \texMathInline{b} in \texMathInline{A}, such that \texMathInline{\forall c \in A : c \not\approx b \Rightarrow b < c}.\relax{\vspace{1ex}
}

\relax{\vspace{1ex}
}\textbf{Definition} (\textit{selection by a complete instantiation{-}based selection rule}) \hspace*{\fill}\\Let \texMathInline{A \in AC\!on\!Atom_{P/\approx}}. Then, the abstract atom selected from A by \texMathInline{<} is the leftmost abstract atom \texMathInline{b}, such that \texMathInline{\forall c \in A : c \not\approx b \Rightarrow b < c}.\relax{\vspace{1ex}
}

In the running example, a complete instantiation{-}based selection rule contains a pair \texMathInline{(ord([g_1,g_2|a_1]),perm(g_1,a_1))}.
There is no aliasing between the elements.
That is, this ordered pair has the same meaning as \texMathInline{(ord([g_1,g_2|a_1]),perm(g_3,a_2))}.

We assume that fully evaluated abstract atoms are dealt with in left{-}to{-}right order.
We also assume that fully evaluated abstract atoms which can be fully evaluated are selected before those which can be unfolded.\NoteBox{\NoteContent{Both assumptions pertaining to fully evaluated atoms are strictly for notational convenience.}}
Under the assumption that more instantiated atoms are always ordered before less instantiated ones, we will represent \texMathInline{<} by its generating set, \texMathInline{Preprior}.
The selection rule \texMathInline{<} itself is inferred from \texMathInline{Preprior} using the assumptions about fully evaluated abstract atoms and the fact that \texMathInline{<} is transitive.
For our running example, formally, \texMathInline{Preprior} contains the pairs \texMathInline{perm(g_1,a_1) < ord([g_1|a_1])} and \texMathInline{ord([g_1,g_2|a_1]) < perm(g_1,a_1)}.

\sectionNewpage

\Ssection{The analysis phase}{The analysis phase}\label{t:x28part_x22analysisx2dphasex22x29}

The analysis consists of a number of abstract derivation trees whose roots are all in a finite set \texMathInline{\mathcal{A}} of abstract conjunctions.
The first of these conjunctions is the abstract conjunction representing the intended call pattern, e.g. \texMathInline{permsort(g_1,a_1)}.
The leaves of these trees are also in \texMathInline{\mathcal{A}}, or they are empty.
Atoms underlined once are selected by the non{-}standard selection rule and an abstraction of resolution is applied to them.
Atoms underlined twice represent atoms which, in a concrete computation, are selected, but an abstraction of resolution is not applied to them.
Instead, these atoms are considered to be completely executed and the result of this execution is collected in a set of output bindings.
That is, the evaluation of such atoms is not interleaved with that of other atoms and we are therefore only interested in the effects they have on the remaining atoms.

\sectionNewpage

\Ssection{A suitable meta{-}interpreter}{A suitable meta{-}interpreter}\label{t:x28part_x22Ax5fsuitablex5fmetax2dinterpreterx22x29}

If each abstract conjunction in \hyperref[t:x28counter_x28x22figurex22_x22permsort1x22x29x29]{Figure~\FigureRef{1}{t:x28counter_x28x22figurex22_x22permsort1x22x29x29}} and \hyperref[t:x28counter_x28x22figurex22_x22permsort2x22x29x29]{Figure~\FigureRef{2}{t:x28counter_x28x22figurex22_x22permsort2x22x29x29}} is assigned the number in front of the conjunction as an identifier and the empty goal is assigned the atom "empty", a simple meta{-}interpreter can run permutation sort under the desired selection rule.

\noindent \begin{SingleColumn}\Scribtexttt{compute(Gs) {\hbox{\texttt{:}}}{-} mi(Gs,1){\hbox{\texttt{.}}}}

\Scribtexttt{mi([],{\char`\_}){\hbox{\texttt{.}}}}

\Scribtexttt{mi([G{\Stttextbar}Gs],State) {\hbox{\texttt{:}}}{-}}

\Scribtexttt{}\mbox{\hphantom{\Scribtexttt{xx}}}\Scribtexttt{selected{\char`\_}index(State,Idx),}

\Scribtexttt{}\mbox{\hphantom{\Scribtexttt{xx}}}\Scribtexttt{divide{\char`\_}goals([G{\Stttextbar}Gs],Idx,Before,Selected,After),}

\Scribtexttt{}\mbox{\hphantom{\Scribtexttt{xx}}}\Scribtexttt{mi{\char`\_}clause(Selected,Body,RuleIdx),}

\Scribtexttt{}\mbox{\hphantom{\Scribtexttt{xx}}}\Scribtexttt{state{\char`\_}transition(State,NewState,RuleIdx),}

\Scribtexttt{}\mbox{\hphantom{\Scribtexttt{xx}}}\Scribtexttt{append(Before,Body,NewGsA),}

\Scribtexttt{}\mbox{\hphantom{\Scribtexttt{xx}}}\Scribtexttt{append(NewGsA,After,NewGs),}

\Scribtexttt{}\mbox{\hphantom{\Scribtexttt{xx}}}\Scribtexttt{mi(NewGs,NewState){\hbox{\texttt{.}}}}

\Scribtexttt{mi([G{\Stttextbar}Gs],State) {\hbox{\texttt{:}}}{-}}

\Scribtexttt{}\mbox{\hphantom{\Scribtexttt{xx}}}\Scribtexttt{selected{\char`\_}index(State,Idx),}

\Scribtexttt{}\mbox{\hphantom{\Scribtexttt{xx}}}\Scribtexttt{divide{\char`\_}goals([G{\Stttextbar}Gs],Idx,Before,Selected,After),}

\Scribtexttt{}\mbox{\hphantom{\Scribtexttt{xx}}}\Scribtexttt{mi{\char`\_}full{\char`\_}eval(Selected,FullAIIdx),}

\Scribtexttt{}\mbox{\hphantom{\Scribtexttt{xx}}}\Scribtexttt{call(Selected),}

\Scribtexttt{}\mbox{\hphantom{\Scribtexttt{xx}}}\Scribtexttt{state{\char`\_}transition(State,NewState,FullAIIdx),}

\Scribtexttt{}\mbox{\hphantom{\Scribtexttt{xx}}}\Scribtexttt{append(Before,After,NewGs),}

\Scribtexttt{}\mbox{\hphantom{\Scribtexttt{xx}}}\Scribtexttt{mi(NewGs,NewState){\hbox{\texttt{.}}}}

\Scribtexttt{divide{\char`\_}goals(Goals,Idx,Before,Selected,After) {\hbox{\texttt{:}}}{-}}

\Scribtexttt{}\mbox{\hphantom{\Scribtexttt{xx}}}\Scribtexttt{length(Before,Idx),}

\Scribtexttt{}\mbox{\hphantom{\Scribtexttt{xx}}}\Scribtexttt{append(Before,[Selected{\Stttextbar}After],Goals){\hbox{\texttt{.}}}}\end{SingleColumn}

Here, \texMathInline{mi/2} is the meta{-}interpretation predicate. It takes a concrete conjunction and a state. The initial call is \texMathInline{compute([permsort(G,A)])} where \texMathInline{G} is instantiated to a ground value and \texMathInline{A} can be any value.
The \texMathInline{mi\_clause/3} predicate encodes clauses as a head, a list of body atoms and a unique identifier for the clause.
The \texMathInline{mi\_full\_eval/2} predicate identifies fully evaluated goals, e.g. \texMathInline{mi\_full\_eval(select(X,Y,Z),fullai1)} to remove an element \texMathInline{X} from a list \texMathInline{Y}, which yields \texMathInline{Z}.
The \texMathInline{selected\_index/2} predicate supplies the index of the atom to be selected in a particular state. The meta{-}interpreter itself does not inspect groundness or aliasing characteristics of conjunctions. Such characteristics are derived during the analysis phase. The \texMathInline{state\_transition/3} predicate expresses which states are directly reachable from which other states and which rules cause the transitions. The full code is in the electronic appendix.

\Ssubsection{Instantiation of the first Futamura projection}{Instantiation of the first Futamura projection}\label{t:x28part_x22Instantiationx5fofx5fthex5ffirstx5fFutamurax5fprojectionx22x29}

Using Logen, the interpreter can be specialized for \texMathInline{mi([permsort(X,Y)],1)}.
We cannot express to Logen that \texMathInline{X} will be instantiated to a ground value, but the control flow is already encoded in \texMathInline{selected\_index/2} and \texMathInline{state\_transition/3}.
Therefore, the result of the specialization will still be a compiled program with the desired control flow.

With regard to the equation \texMathInline{int(s,r)=\alpha(int,s)(r)}, the vector \texMathInline{s} is described in~\cite{futamura_1999_partial} as "a source program and information needed for syntax analysis and semantic analysis". As such, the selection rule is a component of \texMathInline{s} in addition to the source program, \texMathInline{P}. Because the program should work for multiple input queries, a program{'}s top{-}level goal is considered a runtime input.
However, a top{-}level goal instantiated by all runtime top{-}level goals is a static input. Hence, such a goal is a component of \texMathInline{s}, even if \texMathInline{P} is not intended to be run with this goal directly. This is what allows logic programs to be specialized for certain top{-}level calls. Groundness characteristics of the query pattern can also be considered as static information. This is not done in partial deduction of pure logic programs, but we have to take it into account for our approach.

To this end, the compiling control analysis can be encoded as a component of \texMathInline{s}.
Because it provides all required information about the static inputs to the meta{-}interpreter, there is no need to provide those as direct inputs to the evaluator as well.
However, it does \textit{not} necessarily contain enough information about predicates not involved in the coroutining control flow.
That is, certain predicates may be fully evaluated during the analysis phase even if their definition is part of the source program, so the source program is also a component of \texMathInline{s}.
The analysis completely specifies the local and global control for a conjunctive partial deduction of \texMathInline{P} under the semantics implemented by \texMathInline{int}.
In light of the first Futamura projection, an offline partial evaluation of the interpreter must then produce a program equivalent to that produced by the synthesis step~\cite{nys_2017_abstract}.\NoteBox{\NoteContent{We use "partial evaluation" rather than "partial deduction" for impure logic programs.}} In other words, the "classical" synthesis step can be seen as a problem{-}specific shortcut to the outcome of the first Futamura projection.

\sectionNewpage

\Ssection{Programs requiring the \texMathInline{multi} abstraction: primes}{Programs requiring the \texMathInline{multi} abstraction: primes}\label{t:x28part_x22primesx22x29}

The application of compiling control can be recast as an abstract conjunctive partial deduction~\cite{leuschel_2004_framework} in the case of permutation sort.
However, this does not hold for every program with an instantiation{-}based selection rule. This is due to the generation of abstract conjunctions of arbitrary length in some programs. Conjunctive partial deduction can deal with conjunctions of arbitrary length by splitting goals to obtain an \texMathInline{\mathcal{A}}{-}closed set. In our context, however, this would cause a loss of important information regarding aliasing between subconjunctions. The finite analysis of some coroutining programs therefore requires an addition to the abstract domain, known as the \texMathInline{multi} abstraction.

The formal details of this abstraction are quite elaborate. In brief, the concretization of an abstract conjunction containing a \texMathInline{multi} abstraction contains an infinite number of concrete conjunctions of arbitrary length, but with a structure which follows a constrained pattern, expressed as a conjunction.\NoteBox{\NoteContent{In the most recent work on the analysis~\cite{nys_2017_abstract}, we have assumed the pattern of a \texMathInline{multi} to be a conjunction. Known examples of other patterns are artificial. One is provided in the appendix.}} It is easier to illustrate the \texMathInline{multi} through an example first and to indicate the role each component plays rather than to define it beforehand. Here, we will use the following primes generator as an example:

\begin{SingleColumn}\Scribtexttt{primes(N,Primes) {\hbox{\texttt{:}}}{-}}

\Scribtexttt{}\mbox{\hphantom{\Scribtexttt{xx}}}\Scribtexttt{integers(2,I),sift(I,Primes),length(Primes,N)}

\Scribtexttt{integers(N,[]){\hbox{\texttt{.}}}}

\Scribtexttt{integers(N,[N{\Stttextbar}I]) {\hbox{\texttt{:}}}{-} plus(N,1,M),integers(M,I){\hbox{\texttt{.}}}}

\Scribtexttt{sift([N{\Stttextbar}Ints],[N{\Stttextbar}Primes]) {\hbox{\texttt{:}}}{-}}

\Scribtexttt{}\mbox{\hphantom{\Scribtexttt{xx}}}\Scribtexttt{filter(N,Ints,F),sift(F,Primes){\hbox{\texttt{.}}}}

\Scribtexttt{sift([],[]){\hbox{\texttt{.}}}}

\Scribtexttt{filter(N,[M{\Stttextbar}I],F) {\hbox{\texttt{:}}}{-} divides(N,M), filter(N,I,F){\hbox{\texttt{.}}}}

\Scribtexttt{filter(N,[M{\Stttextbar}I],[M{\Stttextbar}F]) {\hbox{\texttt{:}}}{-}}

\Scribtexttt{}\mbox{\hphantom{\Scribtexttt{xx}}}\Scribtexttt{does{\char`\_}not{\char`\_}divide(N,M), filter(N,I,F){\hbox{\texttt{.}}}}

\Scribtexttt{filter(N,[],[]){\hbox{\texttt{.}}}}

\Scribtexttt{length([],0){\hbox{\texttt{.}}}}

\Scribtexttt{length([H{\Stttextbar}T],N) {\hbox{\texttt{:}}}{-} minus(N,1,M),length(T,M){\hbox{\texttt{.}}}}\end{SingleColumn}

This program is run with a top{-}level call \texMathInline{primes(N,P)}, where \texMathInline{N \in \mathbb{N}}.

Rather than list \texMathInline{Preprior} here in full, we will describe it in terms of a set \texMathInline{K\!A = \{integers(g_1,a_1), sift(a_1,a_2), filter(g_1,a_1,a_2), length(a_1,g_1)\}}.
Two rules are sufficient to perform all the comparisons needed to complete the analysis phase:
\texMathInline{\forall x \in K\!A: x \nless integers(g_1,a_1)} and \texMathInline{\forall x \in A\!Atom_{P/\approx}: (\exists y \in K\!A: \gamma(x) \subset \gamma(y)) \Rightarrow \forall y \in K\!A: x < y} (where \texMathInline{\subset} denotes the strict subset relation).

With the resulting order \texMathInline{<} and without further abstraction, the abstract analysis leads to the introduction of abstract conjunctions with an arbitrary number of \texMathInline{filter/3} abstract atoms which are aliased in a consistent manner. Given that \texMathInline{\mathcal{A}} must be finite and that the first argument of \texMathInline{primes/2} can be an arbitrarily large number, adding a conjunction to \texMathInline{\mathcal{A}} for every possible number of filters is not an option. Instead, we abstract away the precise number of filters by replacing the \texMathInline{filter/3} abstract atoms with a \texMathInline{multi} abstraction, which represents any strictly positive natural number of filters.
\texMathInline{multi(filter(g_{id,i,1},a_{id,i,1},a_{id,i,2}),\{a_{id,1,1} = a_1\},\{a_{id,i+1,1} = a_{id,i,2} \},\{ a_{id,\mathcal{L},2} = a_2 \})}, for example, denotes the following set of abstract conjunctions:
\hspace*{\fill}\\
\texMathInline{\{
filter(g_{f_1},a_1,a_2),
filter(g_{f_1},a_1,a_{f_1}) \wedge filter(g_{f_2},a_{f_1},a_2),\hspace*{\fill}\\
filter(g_{f_1},a_1,a_{f_1}) \wedge filter(g_{f_2},a_{f_1},a_{f_2}) \wedge filter(g_{f_3},a_{f_2},a_2),
\ldots \}} where every \texMathInline{f_N} is a unique index which does not occur for that variable type in the conjunction that the set member is part of. The first element of the \texMathInline{multi} abstraction is the conjunctive pattern. In the pattern, \texMathInline{id} represents an identifier unique to this \texMathInline{multi}. This is required as a single abstract conjunction can contain several \texMathInline{multi} abstractions, which need to be distinguished from one another, but whose variables can also be aliased in some cases. The \texMathInline{i} is symbolic and is not used in the pattern itself, but in the remaining arguments to \texMathInline{multi}, which are sets of constaints. The numeric index plays the same part as a regular abstract variable subscript, i.e. to indicate aliasing within an occurrence of the pattern.
The second argument to \texMathInline{multi}, which is called \texMathInline{I\!nit}, is a set of constraints which expresses the aliasing applied to certain variables in the first represented instance of the conjunctive pattern. Here, the symbolic \texMathInline{i} is replaced with \texMathInline{1} to indicate this. The third argument, \texMathInline{Consecutive}, expresses aliasing between consecutive occurrences of the pattern. The fourth, \texMathInline{Final}, expresses aliasing applied to variables in the last represented instance of the pattern. Here, the symbolic \texMathInline{i} is replaced with \texMathInline{\mathcal{L}} (which stands for "last") to indicate this. When an instance of the pattern becomes eligible for abstract resolution, a case split is applied to the \texMathInline{multi} abstraction. The underlying intuition is that it represents either one or multiple occurrences of its pattern. For a more formal and detailed account of the \texMathInline{multi} abstraction, we refer to~\cite{nys_2017_abstract}.

A specification of the control flow using \texMathInline{selected\_index/2} is no longer possible if the \texMathInline{multi} abstraction is required for analysis: in two different concrete instances of an abstract conjunction containing \texMathInline{multi}, the concrete atoms selected for resolution can be at different index positions.
The solution, then, is to transform the program so that this is no longer the case. Transforming the interpreted program \texMathInline{P} into a different purely logical program which can be dealt with {---} at least through a simple transformation {---} is not an option. It is, however, possible to generalize the SLD{-}based execution mechanism implemented by the meta{-}interpreter in a way which still allows the results of partial deduction to be applied. The idea behind this is to introduce an additional operation, \texMathInline{grouping}, which transforms atoms into arguments of an atom with a special predicate symbol, \texMathInline{cmulti}. This operation does not affect the results of the program in any way, but strengthens the correspondence between the abstract analysis and a concrete execution. In this way, the length of \textit{concrete} conjunctions also becomes bounded.
\Ssubsection{An extended meta{-}interpreter}{An extended meta{-}interpreter}\label{t:x28part_x22extx2dmix22x29}

The key modification to the meta{-}interpreter is the addition of the following clauses:

\begin{SingleColumn}\Scribtexttt{mi([G{\Stttextbar}Gs],State) {\hbox{\texttt{:}}}{-}}

\Scribtexttt{}\mbox{\hphantom{\Scribtexttt{xx}}}\Scribtexttt{selected{\char`\_}index(State,Idx),}

\Scribtexttt{}\mbox{\hphantom{\Scribtexttt{xx}}}\Scribtexttt{divide{\char`\_}goals([G{\Stttextbar}Gs],Idx,Before,}

\Scribtexttt{}\mbox{\hphantom{\Scribtexttt{xxxxxxxxxxxxxxx}}}\Scribtexttt{cmulti([building{\char`\_}block(Patt1)]),}

\Scribtexttt{}\mbox{\hphantom{\Scribtexttt{xxxxxxxxxxxxxxx}}}\Scribtexttt{After),}

\Scribtexttt{}\mbox{\hphantom{\Scribtexttt{xx}}}\Scribtexttt{state{\char`\_}transition(State,NewState,one),}

\Scribtexttt{}\mbox{\hphantom{\Scribtexttt{xx}}}\Scribtexttt{append(Patt1,After,NewGsA),}

\Scribtexttt{}\mbox{\hphantom{\Scribtexttt{xx}}}\Scribtexttt{append(Before,NewGsA,NewGs),}

\Scribtexttt{}\mbox{\hphantom{\Scribtexttt{xx}}}\Scribtexttt{mi(NewGs,NewState){\hbox{\texttt{.}}}}

\Scribtexttt{mi([G{\Stttextbar}Gs],State) {\hbox{\texttt{:}}}{-}}

\Scribtexttt{}\mbox{\hphantom{\Scribtexttt{xx}}}\Scribtexttt{selected{\char`\_}index(State,Idx),}

\Scribtexttt{}\mbox{\hphantom{\Scribtexttt{xx}}}\Scribtexttt{divide{\char`\_}goals([G{\Stttextbar}Gs],Idx,Before,}

\Scribtexttt{}\mbox{\hphantom{\Scribtexttt{xxxxxxxxxxxxxxx}}}\Scribtexttt{cmulti([building{\char`\_}block(Patt1),}

\Scribtexttt{}\mbox{\hphantom{\Scribtexttt{xxxxxxxxxxxxxxxxxxxxxxx}}}\Scribtexttt{building{\char`\_}block(Patt2){\Stttextbar}BBs]),}

\Scribtexttt{}\mbox{\hphantom{\Scribtexttt{xxxxxxxxxxxxxxx}}}\Scribtexttt{After),}

\Scribtexttt{}\mbox{\hphantom{\Scribtexttt{xx}}}\Scribtexttt{state{\char`\_}transition(State,NewState,many),}

\Scribtexttt{}\mbox{\hphantom{\Scribtexttt{xx}}}\Scribtexttt{append(Before,Patt1,NewGsA),}

\Scribtexttt{}\mbox{\hphantom{\Scribtexttt{xx}}}\Scribtexttt{append(NewGsA,[cmulti([building{\char`\_}block(Patt2){\Stttextbar}BBs])],NewGsB),}

\Scribtexttt{}\mbox{\hphantom{\Scribtexttt{xx}}}\Scribtexttt{append(NewGsB,After,NewGs),}

\Scribtexttt{}\mbox{\hphantom{\Scribtexttt{xx}}}\Scribtexttt{mi(NewGs,NewState){\hbox{\texttt{.}}}}

\Scribtexttt{mi([G{\Stttextbar}Gs],State) {\hbox{\texttt{:}}}{-}}

\Scribtexttt{}\mbox{\hphantom{\Scribtexttt{xx}}}\Scribtexttt{grouping(State,NextState,Groupings),}

\Scribtexttt{}\mbox{\hphantom{\Scribtexttt{xx}}}\Scribtexttt{apply{\char`\_}groupings([G{\Stttextbar}Gs],Groupings,NewGs),}

\Scribtexttt{}\mbox{\hphantom{\Scribtexttt{xx}}}\Scribtexttt{mi(NewGs,NextState){\hbox{\texttt{.}}}}\end{SingleColumn}

The first two of these clauses extract atoms from a concrete counterpart to a \texMathInline{multi} abstraction. The third one is the one which introduces the concrete counterpart to the \texMathInline{multi} abstraction during interpretation. It uses \texMathInline{apply\_groupings/3} to group certain concrete conjunctions in a concrete \texMathInline{cmulti} structure. The code for this operation is quite long, but it is sufficient to know that the second argument to \texMathInline{apply\_groupings/3} specifies, for a given state, which subconjunctions of the overall goal should be considered instantiations of a \texMathInline{multi} abstraction. This information is available from the abstract interpretation and is encoded using the \texMathInline{grouping/3} predicate, which states which atom indices are grouped during the transition from one state to the next. For example, \texMathInline{grouping(54,55,[[(2,3),(3,4)]])} expresses that, in a concrete instance of the transition from state 54 to state 55, the atoms at index positions 2 (that is, from index position 2 to right before index 3) and 3 (that is, from index position 3 to right before index position 4) instantiate a \texMathInline{multi} and can be grouped together. Each instance of the pattern of the \texMathInline{multi} is further wrapped inside a \texMathInline{building\_block} structure to easily extract a single instance of the pattern from the \texMathInline{cmulti} at a later point in the program. Code for the full meta{-}interpreter and the encoded primes generator, along with instructions on how to generate the analysis, is in the electronic appendix. In the following sections, we show how a specialization of the meta{-}interpreter can be obtained using Logen.

\sectionNewpage

\Ssection{Specialization using Logen}{Specialization using Logen}\label{t:x28part_x22specializationx22x29}

Logen is driven by annotations. Specifically, filter declarations and clause annotations. Filter declarations associate arguments of predicates with so{-}called "binding types". A binding type can be a binding \textit{time}, e.g. "static" or "dynamic", but it can also restrict the structure of the argument. Clause annotations indicate how every call in a clause body should be treated during unfolding. For more detail, we refer to~\cite{leuschel_2002_offline} and~\cite{craig_2004_fully}.

Our filter declarations require the following binding types: "static", "dynamic", "nonvar", "struct" and "list". The first two simply express whether or not an argument will be known at specialization time. The third expresses that an argument has some outermost structure during specialization. The fourth can specify an argument's functor and the binding types of its arguments. The last binding type is for \textit{closed} lists, i.e. lists whose length is bounded. A binding type can also be a disjunction, in which case an argument is considered to be an instance of the first applicable disjunct. The purpose of binding types is this: before a call to a predicate with certain binding types is specialized, the call is generalized and is used as the root of a partial derivation tree. The binding types determine to which extent the call is generalized. A static argument is not generalized, whereas a dynamic one is replaced with a variable.

All clause annotations for the current work can be written using the following strategies: "unfold", "call", "rescall" and "memo". The first means that an atom should be resolved. The second means that a call should be executed at specialization time, while the third means that it should be executed at runtime. The last one means that it should not be unfolded further, but that it should be used as the root for a new derivation tree.

\Ssubsection{Simple meta{-}interpreter}{Simple meta{-}interpreter}\label{t:x28part_x22Simplex5fmetax2dinterpreterx22x29}

For the simple meta{-}interpreter, we start from the naive program annotation performed by Logen. This annotates every predicate call as "unfold" and every builtin call as "call".
We change the recursive call to \texMathInline{mi/2} from "unfold" to "memo", so that it may be specialized separately.
This can be seen as starting a new tree in partial deduction and adding the root of this tree to the set of analyzed conjunctions \texMathInline{\mathcal{A}}.
If the default option, "unfold", is used, new derivation trees are not started and the set of analyzed conjunctions \texMathInline{\mathcal{A}} is not closed.
Furthermore, we annotate the call to \texMathInline{call/1} as "rescall".
Calls to \texMathInline{call/1} apply to atoms which, in the abstract analysis, are underlined twice and which we consider residual.
The "rescall" annotation expresses precisely the idea that a call should simply be executed when the program is run.
The "call" annotation should not be used, as this executes the call during specialization time instead of during program execution.
The most important filters are as follows:

\begin{itemize}\atItemizeStart

\item \begin{SingleColumn}\Scribtexttt{compute(struct(permsort,[dynamic,dynamic]))}\end{SingleColumn}

\noindent We specify as much information about the top{-}level call as possible. We cannot specify that the arguments to the call will have certain groundness characteristics, but this is not necessary as the \texMathInline{selected\_index/2} and \texMathInline{state\_transition/3} predicates are based on this information.

\item \begin{SingleColumn}\Scribtexttt{mi(type(list(nonvar)),static)}\end{SingleColumn}

\noindent Neither the goal, which is represented as a list, nor its elements should be abstracted to variables in the root of a derivation tree. We also keep track of the state index and do not generalize it.
\end{itemize}

The arguments of other filters can be left dynamic: some values \emph{will} be known at specialization time,
but if the calls to an annotated predicate are never used as the root of a derivation tree (and they are not, because they never have the "memo" annotation),
annotating them as "static" has no impact on the partial deduction.

\Ssubsection{Extended meta{-}interpreter}{Extended meta{-}interpreter}\label{t:x28part_x22Extendedx5fmetax2dinterpreterx22x29}

For the extended meta{-}interpreter, the reasoning behind the clause annotations remains mostly the same. However, a crucial change must be applied to deal with the case split on the \texMathInline{multi} abstraction during specialization. During partial evaluation, some information regarding the structure of the \texMathInline{multi} abstraction is necessarily generalized away. To the best of our knowledge, we cannot use filter declarations to preserve info from the abstract analysis about the internal structure of a concrete \texMathInline{multi} \textit{and} generalize calls so that \texMathInline{\mathcal{A}} is closed. The problem is that the specialization procedure must generalize over concrete lists of arbitrary length. This can only be done using a variable, which means that information regarding the contents of the list is lost. Specifically, it becomes impossible to use append operations to build a conjunction at specialization time if some of the appended elements are free variables.
However, we can re{-}encode the required information into the call annotations. We modify and annotate the interpreter as follows.

\begin{SingleColumn}\Scribtexttt{logen(mi/2,mi([B{\Stttextbar}C],A)) {\hbox{\texttt{:}}}{-}}

\Scribtexttt{}\mbox{\hphantom{\Scribtexttt{xxxxxxxx}}}\Scribtexttt{logen(unfold,selected{\char`\_}index(A,D)),}

\Scribtexttt{}\mbox{\hphantom{\Scribtexttt{xxxxxxxx}}}\Scribtexttt{logen(unfold,extracted{\char`\_}patt{\char`\_}one(A,Patt1)),}

\Scribtexttt{}\mbox{\hphantom{\Scribtexttt{xxxxxxxx}}}\Scribtexttt{logen(unfold,}

\Scribtexttt{}\mbox{\hphantom{\Scribtexttt{xxxxxxxxxx}}}\Scribtexttt{divide{\char`\_}goal(}

\Scribtexttt{}\mbox{\hphantom{\Scribtexttt{xxxxxxxxxxxx}}}\Scribtexttt{[B{\Stttextbar}C],D,E,}

\Scribtexttt{}\mbox{\hphantom{\Scribtexttt{xxxxxxxxxxxx}}}\Scribtexttt{cmulti([building{\char`\_}block(Patt1)]),F)),}

\Scribtexttt{}\mbox{\hphantom{\Scribtexttt{xxxxxxxx}}}\Scribtexttt{logen(unfold,state{\char`\_}transition(A,J,one)),}

\Scribtexttt{}\mbox{\hphantom{\Scribtexttt{xxxxxxxx}}}\Scribtexttt{{\hbox{\texttt{.}}}{\hbox{\texttt{.}}}{\hbox{\texttt{.}}}}

\Scribtexttt{}\mbox{\hphantom{\Scribtexttt{xxxxxxxx}}}\Scribtexttt{logen(memo,mi(L,J)){\hbox{\texttt{.}}}}

\Scribtexttt{logen(mi/2,mi([B{\Stttextbar}C],A)) {\hbox{\texttt{:}}}{-}}

\Scribtexttt{}\mbox{\hphantom{\Scribtexttt{xxxxxxxx}}}\Scribtexttt{logen(unfold,selected{\char`\_}index(A,D)),}

\Scribtexttt{}\mbox{\hphantom{\Scribtexttt{xxxxxxxx}}}\Scribtexttt{logen(unfold,}

\Scribtexttt{}\mbox{\hphantom{\Scribtexttt{xxxxxxxxxx}}}\Scribtexttt{extracted{\char`\_}patts{\char`\_}many(A,Patt1,}

\Scribtexttt{}\mbox{\hphantom{\Scribtexttt{xxxxxxxxxxxxxxxxxxxxxxxxxxxxxxx}}}\Scribtexttt{[building{\char`\_}block(Patt2){\Stttextbar}BBs])),}

\Scribtexttt{}\mbox{\hphantom{\Scribtexttt{xxxxxxxx}}}\Scribtexttt{logen(unfold,}

\Scribtexttt{}\mbox{\hphantom{\Scribtexttt{xxxxxxxxxx}}}\Scribtexttt{divide{\char`\_}goal(}

\Scribtexttt{}\mbox{\hphantom{\Scribtexttt{xxxxxxxxxxxx}}}\Scribtexttt{[B{\Stttextbar}C],D,E,}

\Scribtexttt{}\mbox{\hphantom{\Scribtexttt{xxxxxxxxxxxx}}}\Scribtexttt{cmulti([building{\char`\_}block(Patt1),}

\Scribtexttt{}\mbox{\hphantom{\Scribtexttt{xxxxxxxxxxxxxxxxxxxx}}}\Scribtexttt{building{\char`\_}block(Patt2){\Stttextbar}BBs]),F)),}

\Scribtexttt{}\mbox{\hphantom{\Scribtexttt{xxxxxxxx}}}\Scribtexttt{logen(unfold,state{\char`\_}transition(A,M,many)),}

\Scribtexttt{}\mbox{\hphantom{\Scribtexttt{xxxxxxxx}}}\Scribtexttt{{\hbox{\texttt{.}}}{\hbox{\texttt{.}}}{\hbox{\texttt{.}}}}

\Scribtexttt{}\mbox{\hphantom{\Scribtexttt{xxxxxxxx}}}\Scribtexttt{logen(memo,mi(P,M)){\hbox{\texttt{.}}}}\end{SingleColumn}

Here, \texMathInline{extracted\_patt\_one(State,Patt)} defines the structure of the pattern extracted from a \texMathInline{multi} abstraction when a single subconjunction is unfolded. Its counterpart, \texMathInline{extracted\_patts\_many(State,Patt,Patts)}, defines the structure of the extracted pattern and the remaining, non{-}extracted patterns. Neither predicate affects the correctness of the meta{-}interpreter. We have included an annotated version of the meta{-}interpreter without these extra steps in the electronic appendix for comparison.

The filter declarations remain mostly the same. However, goals now consist of program-specific atoms and universal \texMathInline{cmulti/1} atoms. To reflect this, we use the following filter declaration for \texMathInline{mi/2}:

\noindent \begin{SingleColumn}\Scribtexttt{mi(type(}

\Scribtexttt{}\mbox{\hphantom{\Scribtexttt{xxxx}}}\Scribtexttt{list(}

\Scribtexttt{}\mbox{\hphantom{\Scribtexttt{xxxxx}}}\Scribtexttt{(struct(cmulti,}

\Scribtexttt{}\mbox{\hphantom{\Scribtexttt{xxxxxxxxxxxxx}}}\Scribtexttt{[struct({\textquotesingle}{\hbox{\texttt{.}}}{\textquotesingle},}

\Scribtexttt{}\mbox{\hphantom{\Scribtexttt{xxxxxxxxxxxxxxxxxxxxx}}}\Scribtexttt{[struct(building{\char`\_}block,[type(list(nonvar))]),}

\Scribtexttt{}\mbox{\hphantom{\Scribtexttt{xxxxxxxxxxxxxxxxxxxxxx}}}\Scribtexttt{dynamic])]) ;}

\Scribtexttt{}\mbox{\hphantom{\Scribtexttt{xxxxx}}}\Scribtexttt{nonvar))),}

\Scribtexttt{}\mbox{\hphantom{\Scribtexttt{xxx}}}\Scribtexttt{static)}\end{SingleColumn}

\noindent This preserves as much information about each conjunct as possible, as the disjunction (indicated by \Scribtexttt{;}) is not commutative.
While this solution is somewhat ad hoc, it can easily be applied to all examples that we are aware of.
Furthermore, theoretical work on Logen~\cite{craig_2004_fully} mentions the possibility of using custom binding types.
It is possible that a custom binding type for the multi abstraction could help us avoid the above workaround.

\sectionNewpage

\Ssection{Equivalence with the classical approach}{Equivalence with the classical approach}\label{t:x28part_x22synthesisx2dequivalencex22x29}

The programs produced by Logen appear quite different from the syntheses obtained using the classical approach.
To bring them together, we applied the ECCE online specializer to both resulting programs, because it can smooth away some differences through several post{-}processing steps.
The resulting programs produce identical answer sets and display identical finite failure behavior.\footnote{With the exception of the postprocessed classical synthesis of the prime sieve. This suggests a bug in ECCE, but the numbers before postprocessing are still within 16.5\%.}
Furthermore, while they retain some surface differences, they produce their answers after a nearly identical number of inferences.
This demonstrates that both versions implement the same control flow and have nearly identical runtime performance.
For instance, the sum total of the number of inferences required to find all solutions to the 10-queens problem with one approach deviates less than 3.5\% from the number of inferences required with the other approach.
More exhaustive benchmarks can be found in the electronic appendix.
Therefore empirical evidence strongly points to both approaches being equivalent.

The compile{-}time performance for both approaches can safely be said to be nearly identical.
The reason for this is that the analysis phase, which is the most expensive phase, is common to both approaches.

\sectionNewpage

\Ssection{Discussion}{Discussion}\label{t:x28part_x22Discussionx22x29}

We have shown that compiling control can be regarded as a specialization of a specific meta{-}interpreter for logic programs.
This is an application of the first Futamura projection, which answers a question which has long remained open. We have also provided a software pipeline which applies this idea in practice.
We have restricted ourselves here to programs for which depth{-}\texMathInline{k} abstraction is not required, as there is currently no hard{-}and{-}fast rule to ascertain whether depth{-}\texMathInline{k} abstraction affects the eventual program flow. It would be useful to develop such a rule to further extend the set of programs which can be analyzed correctly in an automated way. It is also possible to apply a counterpart to the \texMathInline{multi} abstraction for terms which contain an arbitrary amount of nesting, but which have a predictable, repeating structure. Finally, we plan to investigate a variant of the \texMathInline{multi} abstraction whose pattern is not a conjunction, but a disjunction of conjunctions.

\sectionNewpage

\newpage
\bibliography{bibliography}

\begin{thebibliography}{10}

\bibitem{bruynooghe_1986_compiling}
Bruynooghe, M., Schreye, D.D., Krekels, B.:
\newblock Compiling control.
\newblock In: Proceedings of the 1986 Symposium on Logic Programming, Salt Lake
  City, Utah, USA, September 22-25, 1986, {IEEE-CS} (1986)  70--77

\bibitem{bruynooghe_1989_compiling}
Bruynooghe, M., Schreye, D.D., Krekels, B.:
\newblock Compiling control.
\newblock J. Log. Program. \textbf{6}(1{\&}2) (1989)  135--162

\bibitem{bruynooghe_1991_practical}
Bruynooghe, M.:
\newblock A practical framework for the abstract interpretation of logic
  programs.
\newblock J. Log. Program. \textbf{10}(2) (1991)  91--124

\bibitem{komorowski_1981_specification}
Komorowski, H.J.:
\newblock A specification of an abstract Prolog machine and its application to
  partial evaluation.
\newblock PhD thesis, VTT Grafiska (1981)

\bibitem{gallagher_1986_transforming}
Gallagher, J.P.:
\newblock Transforming logic programs by specialising interpreters.
\newblock In: {ECAI}. (1986)  313--326

\bibitem{nys_2017_abstract}
Nys, V., De~Schreye, D.:
\newblock Abstract conjunctive partial deduction for the analysis and
  compilation of coroutines.
\newblock Formal Aspects of Computing \textbf{29}(1) (Jan 2017)  125--153

\bibitem{leuschel_2004_framework}
Leuschel, M.:
\newblock A framework for the integration of partial evaluation and abstract
  interpretation of logic programs.
\newblock {ACM} Trans. Program. Lang. Syst. \textbf{26}(3) (2004)  413--463

\bibitem{zamalloa_2009_decompilation}
G{\'{o}}mez{-}Zamalloa, M., Albert, E., Puebla, G.:
\newblock Decompilation of java bytecode to prolog by partial evaluation.
\newblock Information {\&} Software Technology \textbf{51}(10) (2009)
  1409--1427

\bibitem{henriksen_2006_abstract}
Henriksen, K.S., Gallagher, J.P.:
\newblock Abstract interpretation of {PIC} programs through logic programming.
\newblock In: Sixth {IEEE} International Workshop on Source Code Analysis and
  Manipulation {(SCAM} 2006), 27-29 September 2006, Philadelphia, Pennsylvania,
  {USA}, {IEEE} Computer Society (2006)  184--196

\bibitem{leuschel_2004_specialising}
Leuschel, M., Craig, S., Bruynooghe, M., Vanhoof, W.:
\newblock Specialising interpreters using offline partial deduction.
\newblock In Bruynooghe, M., Lau, K., eds.: Program Development in
  Computational Logic: {A} Decade of Research Advances in Logic-Based Program
  Development. Volume 3049 of Lecture Notes in Computer Science.
\newblock Springer (2004)  340--375

\bibitem{futamura_1999_partial}
Futamura, Y.:
\newblock Partial evaluation of computation process--an approach to a
  compiler-compiler.
\newblock Higher-Order and Symbolic Computation \textbf{12}(4) (Dec 1999)
  381--391

\bibitem{jones_1993_partial}
Jones, N.D., Gomard, C.K., Sestoft, P.:
\newblock Partial evaluation and automatic program generation.
\newblock Prentice Hall international series in computer science. Prentice Hall
  (1993)

\bibitem{lloyd_1991_partial}
Lloyd, J.W., Shepherdson, J.C.:
\newblock Partial evaluation in logic programming.
\newblock J. Log. Program. \textbf{11}(3{\&}4) (1991)  217--242

\bibitem{leuschel_2002_logic}
Leuschel, M., Bruynooghe, M.:
\newblock Logic program specialisation through partial deduction: Control
  issues.
\newblock {TPLP} \textbf{2}(4-5) (2002)  461--515

\bibitem{leuschel_2002_offline}
Leuschel, M., J{\o}rgensen, J., Vanhoof, W., Bruynooghe, M.:
\newblock Offline specialisation in prolog using a hand-written compiler
  generator.
\newblock CoRR \textbf{cs.PL/0208009} (2002)

\bibitem{craig_2004_fully}
Craig, S., Gallagher, J.P., Leuschel, M., Henriksen, K.S.:
\newblock Fully automatic binding-time analysis for prolog.
\newblock In Etalle, S., ed.: Logic Based Program Synthesis and Transformation,
  14th International Symposium, {LOPSTR} 2004, Verona, Italy, August 26-28,
  2004, Revised Selected Papers. Volume 3573 of Lecture Notes in Computer
  Science., Springer (2004)  53--68

\end{thebibliography}
\postDoc
\end{document}